\documentstyle[aps,preprint]{revtex}

\begin{document}
\title{Stationary Einstein-Maxwell fields in arbitrary dimensions}
\author{Daisuke Ida and Yuki Uchida}
\address{Department of Physics, Tokyo Institute of Technology,
Tokyo 152-8551, Japan}
\date{\today}
\maketitle
\begin{abstract}
The Einstein-Maxwell equations in $D$-dimensions admitting
$(D-3)$ commuting Killing vector fields have been investigated.
The existence of the electric, magnetic and twist potentials have been proved.
The system is formulated as the harmonic map coupled to gravity on three-dimensional base space generalizing
the Ernst system in the four-dimensional stationary Einstein-Maxwell theory.
Some classes of the new exact solutions have been provided, which include
the electro-magnetic generalization of the 
Myers-Perry solution, which describes the rotating black hole immersed
in a magnetic universe, and the static charged black ring solution.

\end{abstract}
\section{Introduction}

The theory of the Einstein-Maxwell equations has been extensively developed in
conventional four-dimensional general relativity.
One remarkable goal of this theory is the uniqueness theorem of the
Kerr-Newman {\em et al.} solutions\cite{KN} among the stationary charged black hole solutions\cite{Mazur,Bunting}.

On the other hand, there is much interest in the black holes in higher dimensional
theories\cite{MP}, in particular, in the context of the TeV scale gravity.
The TeV scale gravity is realized in the
large extra dimension scenario\cite{ADD} or in the warped compactification scenario\cite{RS},
where the fundamental Planck scale in higher dimensions can be set around TeV.
As pointed out in Refs.~\cite{BF,GT,DL}, higher dimensional black holes might be produced at 
near future colliders such as the CERN Large Hadron Collider (LHC).
These black hole production events have a very important meaning in physics,
since we might be able to prove the existence of the extra dimensions 
and we may 
get some information on the quantum gravity in an experimental method.
Therefore, it is significant to clarify the elementary physical processes
in black hole productions at colliders.
The black holes so produced will have quadru or higher multipole moments, so that
they emit these multipole moments in the form of the gravitational radiation 
to settle down to the stationary black hole state. This process is called the balding phase
in the literature.
From a point of view of phenomenology, 
one question is how much the black hole emits its mass and angular momenta 
in this balding phase.
Another question is as to the stationary state of the black holes.
By analogy with the four-dimensional case, the end product of the balding phase
is assumed to be described by the Myers-Perry (higher-dimensional Kerr) 
black hole \cite{MP}.
The black hole uniqueness theorem however has not yet been established in higher dimensions
except for the static case\cite{unique1,unique2}. 
Since the differential cross section of rotating black hole is larger for 
larger angular momentum, the static approximation will be no more good in general\cite{Ida1}.
Hence, we need to study more on the stationary black holes.

We should also take account of the effect of the black hole charge,
because the black hole might be produced with the electric charge 
and even the neutral black hole may be charged in the evaporation phase after
the balding phase by the Hawking emission\cite{Hawking} of the charged particles.
One way to consider the charged black hole might be the investigation of the
low energy effective theory of the string, in which case
we have charged black hole solutions.
Another way is to consider the more simple and conventional Einstein-Maxwell system.
The charged black holes in Einstein-Maxwell system might be quite different from those
in the string theory\cite{HS}, since in the latter cases, the black hole
carries the massless scalar fields, which are regarded as being stabilized in some way.

In four dimensional general relativity, the asymptotically flat,
stationary charged black holes in Einstein-Maxwell theory
is uniquely determined by the Kerr-Newman {\em et al.} solution\cite{Mazur,Bunting,Carter}.
The stationary charged black hole solution has not yet been known for the Einstein-Maxwell
system expect for the static Reissner-Nordstroem solution or the Majumdar-Papapetrou
multi black hole solutions\cite{Ma,Pa} describing the equilibrium state of many extremely charged
 black holes. (The stationary extremely charged black hole solution is 
 known for the five-dimensional Einstein-Maxwell system with 
 Chern-Simons term \cite{GMT}.) 
We also need non-asymptotically flat charged black hole solutions,
since the higher dimensional generalization of the Kerr-Newman {\em et al.} black hole solution
might not be appropriate for describing the black holes produced at colliders.
To describe the gravitational field in the neighborhood of 
a neutral black hole at LHC, the asymptotically flat Myers-Perry solution gives a
good approximation, because the effect of the brane tension can be estimated
to be negligible at the event horizon for such small enough black holes.
For charged black holes, the existence of the brane has an effect on the electro-magnetic
field, though its gravitational field can be negligible. 
This is just the mechanism of the brane world scenario; the standard model fields are confined
to the brane and only the gravitation propagates in the bulk,
by which the fundamental scale can be set to that low energy\cite{ADD,RS}.
Thus, the charged black holes on brane is such that the electric fields associated with
the black hole charge is distributed only on the three-brane.
This means that, even when we consider static charged black holes,
the spherically symmetric (in higher dimensional sense) 
Reissner-Nordstroem solution would not provide a good description
of a black hole on the brane.

From the purely theoretical point of view, 
whether the higher-dimensional generalization of the Kerr-Newman {\em et al.} solution exists,
or whether it gives the unique asymptotically flat stationary charged black hole solution
in Einstein-Maxwell system is an interesting question.
The Ernst formulation\cite{Ernst} of the stationary-axisymmetric Einstein-Maxwell
field equations is useful for both problems in four-dimensional general relativity\cite{Carter}.
In this formalism, the Einstein-Maxwell fields are described by a pair of complex scalar
fields, and the system is described by a kind of nonlinear $\sigma$-model (harmonic map)
on the two dimensional base space. 
This reduction relies on the existence of the twist potential, by which the 
twist of the Killing vector field, which is originally a vector like object, turns out
to be essentially a scalar field.
 To proceed the black hole uniqueness theorem in higher dimensions,
formulation of the boundary value problem for black hole solutions
\cite{Carter} will be a necessary step.

The subject of this paper is the Einstein-Maxwell field with Killing vector fields
in arbitrary dimensions, which reduce to a harmonic map coupled to gravity.
The formulation presented here will be useful to construct the charged black hole solutions
and it might be helpful to formulate the boundary value problem for the black hole
solutions.

In Sec.~\ref{II},
the Einstein-Maxwell equations in $D$-dimensional space-time with $(D-3)$ commuting
Killing vector fields are considered.
We show the existence of the twist potential extending the result of 
Maison \cite{Dvac} 
for the vacuum case, and that the system reduces to a harmonic map coupled to gravity.
In Sec.~\ref{III}, we assume another commuting Killing vector field.
Then we
show that the metric can be written in block diagonal form under appropriate
conditions, and the system 
reduce to the harmonic map on the two-dimensional base space.
In Sec.~\ref{IV}, examples of the five-dimensional Einstein-Maxwell fields are given.
We show an electro-magnetic generalization of the Myers-Perry black hole, which represents
the rotating black hole immersed in a magnetic field.
In Sec.~\ref{V}, we consider an electro-magnetic generalization of the 
Emparan-Reall static solutions in arbitrary dimensions. We give the
charged generalization of the static black ring solution.
We summarize in Sec.~\ref{VI}.

\section{Einstein-Maxwell fields admitting Killing vector fields}
\label{II}
We consider the $D$-dimensional Einstein-Maxwell system given by the Lagrangian
\begin{equation}
{\cal L}=R-F^2,
\end{equation}
generalizing the formulation given by Maison\cite{Dvac} in vacuum case.
Assume that the space-time admits $(D-3)$ 
commuting Killing vector fields $\xi_I=\partial_I$ $(I=4,\cdots,D)$,
such that ${\mbox \pounds}_{\xi_I}g=0$, $[\xi_I,\xi_J]=0$.
Then, the metric can be put in the form
\begin{eqnarray}
g=|f|^{-1}\gamma_{ij}dx^idx^j+f_{IJ}(dx^I+w^I{}_idx^i)(dx^J+w^J{}_jdx^j),
\end{eqnarray}
where $i,j=1,2,3$, $f={\rm det}(f_{IJ})$
and $\gamma_{ij}$, $w^I{}_i$ and the {\em gravitational potential} $f_{IJ}$ are independent on
coordinates $x^I$.
We define the {\em twist 1-form} $\omega_I$ by
\begin{eqnarray}
\omega_{I\mu}=|f|f_{IJ}\sqrt{|\gamma|}\epsilon_{ij\mu}\gamma^{im}
\gamma^{jn}\partial_mw^J{}_n
\label{defomega}
\end{eqnarray}
where $\mu=1,\cdots,D$,
$\gamma={\rm det}(\gamma_{ij})$, $\gamma^{ij}$ is the inverse metric of $\gamma_{ij}$
and $\epsilon_{ijk}$ denotes the totally skew-symmetric symbol such that $\epsilon_{123}=1$,
 $\epsilon_{I\mu\nu}=0$.
Using the Killing condition $\nabla_\mu\xi_{I\nu}=\nabla_{[\mu}\xi_{I\nu]}$ and
the commutation relation
\begin{equation}
[\xi_I,\xi_J]^\mu=
\xi_I{}^\nu\nabla_\nu \xi_J{}^\mu-\xi_J{}^\nu\nabla_\nu \xi_I{}^\mu
\end{equation}
One can prove the following formula
\begin{equation}
\partial_\mu f_{IJ}=2\xi_I{}^\nu \partial_\mu \xi_{J\nu}
\label{df}
\end{equation}
Then we obtain the useful expression
\begin{eqnarray}
2\partial_{[\mu}\xi_{I\nu]}
=-2f^{JK}\xi_{J[\mu}\partial_{\nu]}f_{KI}
-f^{-1}\sqrt{|\gamma|}\gamma^{mn}\epsilon_{\mu\nu m}\omega_{In},
\label{dxi}
\end{eqnarray}
where 
$f^{IJ}$ denotes the inverse matrix of $f_{IJ}$.
Taking the divergence of Eq.~(\ref{df}),
we obtain the Laplacian of $f_{IJ}$, which gives the $(I,J)$-component of the Ricci tensor:
\begin{eqnarray}
\nabla^2 f_{IJ}&=&2(\nabla_\mu\xi_{I\nu})(\nabla^\mu\xi_J{}^\nu)
+2\xi_I{}^\nu\nabla^2\xi_{J\nu}\nonumber\\
&=&f^{KL}f_{KI,\mu} f_{LJ}{}^{,\mu}-f^{-1}\omega_{I\mu}\omega_J{}^\mu
-2R_{IJ},
\end{eqnarray}
where we have used Eq.~(\ref{dxi}) and the identity for the Killing vector field
$\nabla^2\xi_{I\mu}=-R_{I\mu}$.
The three-dimensional curl of Eq.~(\ref{defomega}) gives the $(I,i)$-component of the Ricci tensor:
\begin{eqnarray}
2\partial_{[i}\omega_{Ij]}=
2|f|^{-1}\sqrt{|\gamma|}\epsilon_{ijk}R_I^k.
\label{domega}
\end{eqnarray}
On the other hand, the three-dimensional divergence of Eq.~(\ref{defomega}) gives
\begin{equation}
D\cdot\omega_I=f^{-1}Df\cdot \omega_I+f^{JK}Df_{IJ}\cdot \omega_K,
\label{divomega}
\end{equation}
where $D$ is the linear connection compatible with the three-metric $\gamma_{ij}$
and the dot denotes the inner product naturally defined by $\gamma_{ij}$.
The $(i,j)$-component of the Ricci tensor is obtained by calculating the Ricci tensor
with respect to the three-metric $\gamma_{ij}$:
\begin{eqnarray}
{}^{(\gamma)}R_{ij}&=&
f^{-2}\gamma_{im}\gamma_{jn}R^{mn}
+|f|f^{IJ}R_{IJ}\gamma_{ij}\nonumber\\
&&{}+{1\over 4}f^{-2}f_{,i}f_{,j}
+{1\over 4}f^{IJ}f^{KL}f_{IK,i}f_{JL,j}
+{1\over 2}f^{-2}f^{IJ}\omega_{Ii}\omega_{Jj}.
\end{eqnarray}

Next, we consider the Maxwell field $F=dA$.
Assume that there is a gauge such that the $U(1)$ gauge field respects the space-time symmetry
${\mbox\pounds}_{\xi_I}A=0$. 
Then we have
\begin{equation}
F_{IJ}=2\partial_{[I}A_{J]}=0.
\end{equation}
Let us define the {\em electric 1-form} $E_I$ and {\em magnetic 1-form} $B$ by
\begin{eqnarray}
E_{I\mu}&=&-F_{I\mu},
\label{defE}\\
B_{\mu}&=&
{1\over 2}|f|\sqrt{|\gamma|}\epsilon_{mn\mu}F^{mn}
\label{defB}\\
&=&{1\over 2}|f|\sqrt{|\gamma|}\epsilon_{mn\mu}\gamma^{mp}\gamma^{nq}(F_{pq}+2F_{Ip}w^I{}_q).
\label{defB2}
\end{eqnarray}
These are three-dimensional 1-forms in the sense $\xi_I{}^\mu E_{J\mu}=\xi_I{}^\mu B_\mu=0$.
The Maxwell equation $\partial_{[\mu}F_{\nu I]}=0$ gives
\begin{equation}
\partial_{[\mu}E_{I\nu]}=0.
\end{equation}
Therefore, there locally exist {\em electric potentials} $\Phi_I$ such that
\begin{equation}
\partial_{\mu}\Phi_I=E_{I\mu}.
\end{equation}
The exterior differentiation of
Eq.~(\ref{defB}) with the Maxwell equation $\partial_m (\sqrt{-g}F^{mn})=0$ gives
\begin{equation}
\partial_{[\mu}B_{\nu]}=0,
\end{equation}
which ensures the local existence of {\em magnetic potential} $\Psi$ such that
\begin{equation}
\partial_{\mu}\Psi=B_\mu.
\end{equation}
Then, the field strength can be written in the form
\begin{equation}
F_{\mu\nu}=-2f^{IJ}\xi_{I[\mu}\partial_{\nu]}\Phi_J
-f^{-1}\sqrt{|\gamma|}\gamma^{mn}\epsilon_{\mu\nu m}\partial_n\Psi.
\label{fieldstrength}
\end{equation}
Taking the divergence of Eq.~(\ref{defE}) and using $\partial_\mu F^{\mu\nu}=0$,
we obtain
\begin{equation}
\nabla^2 \Phi_I=F_{\mu\nu}\nabla^\mu\xi_I{}^\nu
=f^{JK}f_{JI,\mu}\Phi_K{}^{,\mu}-f^{-1}\Psi_{,\mu}\omega_I{}^\mu
\end{equation}
On the other hand, the three-dimensional divergence of Eq.~(\ref{defB2}) gives
\begin{equation}
D^2\Psi=f^{-1}Df\cdot D\Psi+f^{IJ}D\Phi_I\cdot\omega_J.
\end{equation}
The stress-energy tensor of the Maxwell field is given by
\begin{equation}
T_{\mu\nu}=F_{\mu\lambda}F_\nu{}^\lambda-{1\over 4}F_{\lambda\sigma}F^{\lambda\sigma}g_{\mu\nu}.
\end{equation}
Then, by substituting Eq.~(\ref{fieldstrength}) into  the Einstein equation 
$R_{\mu\nu}=T_{\mu\nu}-(D-2)^{-1}Tg_{\mu\nu}$, we obtain
\begin{eqnarray}
R_{\mu\nu}&=&
\left[
f^{IK}f^{JL}\Phi_{I,\lambda}\Phi_J{}^{,\lambda}
-{1\over D-2}(
f^{IJ}f^{KL}\Phi_{I,\lambda}\Phi_J{}^{,\lambda}
-f^{-1}f^{KL}\Psi_{,\lambda}\Psi^{,\lambda})
\right]\xi_{K\mu}\xi_{L\nu}
\nonumber\\
&&{}+2f^{-1}f^{IJ}\xi_{I(\mu}S_{J\nu)}
+f^{IJ}\Phi_{I,\mu}\Phi_{J,\nu}+f^{-1}\Psi_{,\mu}\Psi_{,\nu}
\nonumber\\
&&{}
-{1\over D-2}\left[
f^{IJ}\Phi_{I,\lambda}\Phi_J{}^{,\lambda}
+(D-3)f^{-1}\Psi_{,\lambda}\Psi^{,\lambda}
\right]|f|^{-1}\gamma_{\mu\nu},
\end{eqnarray}
where we have defined the {\em Poynting 1-form} $S_I$ by
\begin{equation}
S_{I\mu}=|f|\sqrt{|\gamma|}\epsilon_{ij\mu}\gamma^{im}\gamma^{jn}\Phi_{I,m}\Psi_{,n}.
\end{equation}
Then, Eq.~(\ref{domega}) becomes
\begin{equation}
\partial_{[\mu}\omega_{I\nu]}=-2\partial_{[\mu}\Phi_I\partial_{\nu]}\Psi,
\label{domega2}
\end{equation}
which implies the local existence of the {\em twist potential} $\lambda_I$, such that
\begin{equation}
\partial_\mu\lambda_I=\omega_{I\mu}+\Phi_I\partial_\mu\Psi-\Psi\partial_\mu\Phi_I.
\end{equation}

Hence, the configuration of the Einstein-Maxwell fields is encoded in 
the $(D-3)(D-2)/2$ gravitational potentials $f_{IJ}$, $(D-3)$ twist potentials $\lambda_I$,
$(D-3)$ electric potentials $\Phi_I$ and a magnetic potential $\Psi$ 
on the three-dimensional space with the metric $\gamma_{ij}$.
The complete set of the field equations are given in three-dimensional form by
\begin{eqnarray}
D^2f_{IJ}&=&f^{KL}Df_{IK}\cdot Df_{JL}
-f^{-1}(D\lambda_I+\Psi D\Phi_I-\Phi_I D\Psi)\cdot(D\lambda_J+\Psi D\Phi_J-\Phi_J D\Psi)
\nonumber\\
&&{}
-2D\Phi_I\cdot D\Phi_J
+{2\over D-2}f_{IJ}(f^{KL}D\Phi_K\cdot D\Phi_L-f^{-1}D\Psi\cdot D\Psi),
\label{laplacianf}\\
D^2\lambda_I&=&
f^{-1}(Df+\Psi D\Psi)\cdot(D\lambda_I+\Psi D\Phi_I-\Phi_I D\Psi)
\nonumber\\
&&{}
+f^{JK}(Df_{IJ}+\Phi_I D\Phi_J)\cdot (D\lambda_K+\Psi D\Phi_K-\Phi_K D\Psi)
\nonumber\\
&&{}
+(f^{-1}\Phi_I Df\cdot D\Psi-f^{JK}\Psi Df_{IJ}\cdot D\Phi_K)
\label{laplacianlambda},\\
D^2\Phi_I&=&f^{JK}Df_{IJ}\cdot D\Phi_K
-f^{-1}D\Psi\cdot (D\lambda_I+\Psi D\Phi_I-\Phi_I D\Psi)
\label{laplacianphi},\\
D^2\Psi&=&f^{-1}Df\cdot D\Psi
+f^{IJ}D\Phi_I\cdot (D\lambda_J+\Psi D\Phi_J-\Phi_J D\Psi)
\label{laplacianpsi},
\end{eqnarray}
and by the  Einstein equation on the three-space
\begin{eqnarray}
{}^{(\gamma)}R_{mn}&=&
f^{IJ}\Phi_{I,m}\Phi_{J,n}
+f^{-1}\Psi_{,m}\Psi_{,n}
+{1\over 4}f^{-2}f_{,m}f_{,n}
+{1\over 4}f^{IJ}f^{KL}f_{IK,m}f_{JL,n}\nonumber\\
&&{}+{1\over 2}f^{-1}f^{IJ}
(\lambda_{I,m}+\Psi \Phi_{I,m}-\Phi_I \Psi_{,m})
(\lambda_{J,n}+\Psi \Phi_{J,n}-\Phi_J \Psi_{,n}).
\label{3Ricci}
\end{eqnarray}
The system is then described by the action
\begin{eqnarray}
S&=&\int dx^3 \sqrt{|\gamma|}\left[
{}^{(\gamma)}R
-f^{IJ}\partial \Phi_I\cdot \partial\Phi_J
-f^{-1}(\partial\Psi)^2
-{1\over 4}f^{-2}(\partial f)^2
-{1\over 4}f^{IJ}f^{KL}\partial f_{IK}\cdot \partial f_{JL}
\right.
\nonumber\\
&&{}
\left.-{1\over 2}f^{-1}f^{IJ}
(\partial\lambda_I+\Psi\partial\Phi_I-\Phi_I \partial\Psi)
\cdot (\partial\lambda_J+\Psi\partial\Phi_J-\Phi_J \partial\Psi)
\right],
\end{eqnarray}
of the harmonic map coupled to gravity. 
The metric function $w^I{}_i$ is obtained by solving
\begin{equation}
2\partial_{[i}w^I{}_{j]}=-f^{-1}f^{IJ}\sqrt{|\gamma|}\epsilon_{ijm}
\gamma^{mn}(\lambda_{J,n}+\Psi \Phi_{J,n}-\Phi_J \Psi_{,n}).
\end{equation}
The integrability condition of this equation has been given by Eq.~(\ref{divomega})
or Eqs.~(\ref{laplacianlambda}), (\ref{laplacianphi}) and (\ref{laplacianpsi}).
\section{Weyl-Papapetrou--type metrics}
\label{III}
Here we assume the existence of another Killing vector field $\xi_3=\partial_3$ 
which commutes with
the other Killing vectors $[\xi_3,\xi_I]=0$.
The gauge field is also assumed to be independent of the coordinate $x^3$.
Then we consider the condition 
 that the two-dimensional distribution orthogonal to $\forall\xi_{\tilde I}$ 
($\tilde I=3,\cdots,D$) is integrable, namely surface forming.
This integrability condition is given by the Frobenius condition
\begin{eqnarray}
\xi_3{}^{[\mu_3}\cdots\xi_D{}^{\mu_D}\partial^\nu\xi_{\tilde I}{}^{\lambda]}
=0.~~~(\tilde I=3,\cdots,D)
\label{frobenius}
\end{eqnarray}
We note that the twist 1-form can be written also in the form
\begin{equation}
\omega_{I\mu}=\sqrt{-g}\epsilon_{\mu\nu\lambda\mu_4\cdots\mu_D}
\xi_4{}^{\mu_4}\cdots\xi_D{}^{\mu_D}\partial^\nu\xi_I{}^\lambda,
\end{equation}
where $\epsilon_{\mu_1\cdots\mu_D}$ is the totally skew-symmetric symbol 
$\epsilon_{1\cdots D}=1$.
Then, the condition~(\ref{frobenius}) for $\tilde I\ne 3$ turns out to be equivalent with
\begin{equation}
\omega_{I3}=0.
\label{omega3}
\end{equation}
By differentiating  Eq.~(\ref{omega3}) and using  Eq.~(\ref{domega2}), we obtain
\begin{eqnarray}
\partial_\mu\omega_{I3}
=-2\partial_{[3}\omega_{I\nu]}
=4\partial_{[3}\Phi_I\partial_{\nu]}\Psi=0.
\end{eqnarray}
Therefore, the Frobenius condition for $\tilde I\ne 3$ is satisfied if 
it holds somewhere at a point, which we assume in what follows.
Since the geometric assumption on $\xi_3$ and $\xi_I$ is symmetric, 
we assume that the Frobenius condition is satisfied for all $\tilde I$.
This is just a simplifying assumption at this point.
In that case, $w^I{}_1=w^I{}_2=0$ and the three-dimensional
metric can be written in the form
\begin{eqnarray}
\gamma_{ij}&=&\hat\gamma_{pq}dx^p dx^q+\epsilon_3\rho^2 d\eta^2,
\end{eqnarray}
where $p,q=1,2$, $\eta=x^3$ and $\epsilon_3=\pm1$ according to the sign of $f$.
The non-zero components of the Ricci tensor with respect to $\gamma_{ij}$ are
\begin{eqnarray}
{}^{(\gamma)}R_{\eta\eta}&=&-\epsilon_3\rho\hat D^2\rho,
\label{R00}\\
{}^{(\gamma)}R_{pq}&=&K\hat \gamma_{pq}-{1\over\rho}\hat D_p\hat D_q\rho
\label{Rpq},
\end{eqnarray}
where $\hat D$ and $K$ denotes
 the linear connection 
and the Gaussian curvature on the two-dimensional space with the metric $\hat\gamma_{pq}$.
Since from Eq.~(\ref{3Ricci}) $R_{\eta\eta}=0$ holds, 
Eq.~(\ref{R00}) implies that the function $\rho$ is a harmonic function in the two-dimensional
orthogonal surface.
Let us take the isothermal coordinates $x^1=\rho$ and $x^2=z$, such that
$\hat D^2 z=0$, $\hat D_p\rho\hat D^p z=0$ and 
$\hat D_p z\hat D^p z=\hat D_p\rho\hat D^p \rho$.
Then, $\hat\gamma_{pq}$ takes a conformally flat form and the metric
can be written in the Weyl-Papapetrou--type form\cite{Papapetrou}
\begin{eqnarray}
g=|f|^{-1}e^{2\sigma}(d\rho^2+dz^2)-f^{-1}\rho^2d\eta^2
+f_{IJ}(dx^I+w^I d\eta)(dx^J+w^Jd\eta),
\label{weyl-papapetrou}
\end{eqnarray}
where all the metric function depends only on $\rho$ and $z$.
Then, the Einstein-Maxwell fields are given by axisymmetric solution of 
Eqs.~(\ref{laplacianf}), (\ref{laplacianlambda}), (\ref{laplacianphi}) and 
(\ref{laplacianpsi}) on the abstract flat three-space with the metric
\begin{eqnarray}
\tilde\gamma=d\rho^2+dz^2+\rho^2 d\varphi^2,
\end{eqnarray}
in the usual cylindrical coordinates.
For example, 
$D^2 \phi$ and
$D\phi\cdot D\psi$
are regarded as
$\phi_{,\rho\rho}+\rho^{-1}\phi_{,\rho}+\phi_{,zz}$
and
$\phi_{,\rho}\psi_{,\rho}+\phi_{,z}\psi_{,z}$, respectively.
Once the potentials $f_{IJ}$, $\lambda_I$, $\Phi_I$ and $\Psi$ are
obtained, the Eq.~(\ref{3Ricci}) gives the gradient of the metric function $\sigma$:
\begin{eqnarray}
{2\over\rho}\sigma_{,\rho}
&=&{}^{(\gamma)}R_{\rho\rho}-{}^{(\gamma)}R_{zz}\nonumber\\
&=&f^{IJ}(\Phi_{I,\rho}\Phi_{J,\rho}-\Phi_{I,z}\Phi_{J,z})
+f^{-1}[(\Psi_{,\rho})^2-(\Psi_{,z})^2]
\nonumber\\
&&{}+{1\over 4}f^{-2}[(f_{,\rho})^2-(f_{,z})^2]
+{1\over 4}f^{IJ}f^{MN}(f_{IM,\rho}f_{JN,\rho}-f_{IM,z}f_{JN,z})
\nonumber\\
&&{}+{1\over 2}f^{-1}f^{IJ}
\left[\left(\lambda_{I,\rho}+\Psi\Phi_{I,\rho}-\Phi_I\Psi_{,\rho}\right)
\left(\lambda_{J,\rho}+\Psi\Phi_{J,\rho}-\Phi_J\Psi_{,\rho}\right)\right.
\nonumber\\
&&{}\left.-
\left(\lambda_{I,z}+\Psi\Phi_{I,z}-\Phi_I\Psi_{,z}\right)
\left(\lambda_{J,z}+\Psi\Phi_{J,z}-\Phi_J\Psi_{,z}\right)\right],\\
\label{sigmarho}
{1\over\rho}\sigma_{,z}&=&
{}^{(\gamma)}R_{\rho z}\nonumber\\
&=&
f^{IJ}\Phi_{I,\rho}\Phi_{J,z}
+f^{-1}\Psi_{,\rho}\Psi_{,z}
\nonumber\\
&&{}+{1\over 4}f^{-2}f_{,\rho}f_{,z}
+{1\over 4}f^{IJ}f^{MN}f_{IM,\rho}f_{JN,z}
\nonumber\\
&&{}+{1\over 2}f^{-1}f^{IJ}
\left(\lambda_{I,\rho}+\Psi\Phi_{I,\rho}-\Phi_I\Psi_{,\rho}\right)
\left(\lambda_{J,z}+\Psi\Phi_{J,z}-\Phi_J\Psi_{,z}\right).
\label{sigmaz}
\end{eqnarray}
The integrability condition $\sigma_{,\rho z}=\sigma_{,z\rho}$ of Eqs.~(\ref{sigmarho}) and (\ref{sigmaz})
is assured by the field equations~(\ref{laplacianf}),
The metric function $w^I$ is then obtained by solving
\begin{eqnarray}
w^I{}_{,\rho}&=&|\rho| f^{-1}f^{IJ}
(\lambda_{J,z}+\Psi \Phi_{J,z}-\Phi_J \Psi_{,z}),\\
w^I{}_{,z}&=&-|\rho| f^{-1}f^{IJ}
(\lambda_{J,z}+\Psi \Phi_{J,z}-\Phi_J \Psi_{,z}).
\end{eqnarray}
\section{Einstein-Maxwell fields in five dimensions}
\label{IV}
Let us investigate the five-dimensional Einstein-Maxwell fields
admitting two commuting Killing vector fields.
To simplify the problem, we seek for the solution in the form
$f_{44}=u$,
$f_{55}=v$,
$\lambda_5=\lambda$,
$\Phi_4=\Phi$,
$f_{45}=\lambda_4=\Phi_5=\Psi=0$.
Then, Eqs.~(\ref{laplacianf}),
become
\begin{eqnarray}
D^2u&=&{Du\cdot Du\over u}
-{4D\Phi\cdot D\Phi\over 3},
\label{ddu}
\\
D^2v&=&{Dv\cdot Dv\over v}
-{D\lambda\cdot D\lambda\over uv}
+{2vD\Phi\cdot D\Phi\over 3u},
\label{ddv}
\\
D^2\lambda&=&
{Du\cdot D\lambda\over u}
+{2Dv\cdot D\lambda\over v},
\label{ddlambda}\\
D^2\Phi&=&{Du\cdot D\Phi\over u}.
\label{ddphi}
\end{eqnarray}
The vacuum case ($\Phi=0$) was previously considered by Mazur and 
Bombelli in the context of the Kaluza-Klein black holes\cite{KKBH}.
Firstly, we solve Eqs.~(\ref{ddu}) and (\ref{ddphi}).
We assume that the variables $u$ and $\Phi$ are functions of the
harmonic function $\phi$; $u=u(\phi)$, $\Phi=\Phi(\phi)$, such that 
\begin{equation}
D^2\phi=0.
\label{harmonic}
\end{equation}
Then Eqs.~(\ref{ddu}) and (\ref{ddphi}) become
\begin{eqnarray}
u''&=&{(u')^2\over u}-{4\over 3}(\Phi')^2,
\label{difu}\\
\Phi''&=&{u'\over u}\Phi'.
\label{difphi}
\end{eqnarray}
The general solution is
\begin{eqnarray}
u&=&
{4\epsilon_4 C_3{}^2 
\over
(e^{C_1\phi+C_2}+\epsilon_4e^{-C_1\phi-C_2})^2},\\
\Phi&=&\sqrt{3\over 2}C_3
{e^{C_1\phi+C_2}-\epsilon_4e^{-C_1\phi-C_2}
\over
e^{C_1\phi+C_2}+\epsilon_4e^{-C_1\phi-C_2}}
+C_4,
\end{eqnarray}
where $\epsilon_4=\pm 1$. 
Note that integration constants $C_1,\cdots,C_4$ can be absorbed 
into the definition of $\phi$,
the scaling of the coordinates $x^4$, $x^5$ and the gauge transformation.
Next, if we put $v=w(\phi)\kappa$, then Eqs.~(\ref{ddv}) and (\ref{ddlambda}) become
\begin{eqnarray}
D^2\kappa&=&{(D\kappa)^2\over\kappa}-{(D\lambda)^2\over\kappa}(uw^2)^{-1}
-\left[{w''\over w}-{(w')^2\over w^2}-{2(\Phi')^2\over 3 u}\right]\kappa (D\phi)^2,
\label{ddkappa2}
\\
D^2\lambda&=&
{2D\kappa\cdot D\lambda\over\kappa}+\left({u'\over u}+{2w'\over w}\right)D\phi\cdot d\lambda.
\label{ddlambda2}
\end{eqnarray}
If we choose $w=|u|^{-1/2}$, then, by virtue of Eq.~(\ref{difu}), the third term of the right hand side (r.h.s.) of
Eq.~(\ref{ddkappa2}) becomes zero,
and the second term of r.h.s. of Eq.~(\ref{ddlambda2}) also vanishes.
Then, Eqs.~(\ref{ddkappa2}) and (\ref{ddlambda2}) become
\begin{eqnarray}
D^2\kappa&=&{(D\kappa)^2-\epsilon(D\lambda)^2\over\kappa}
\label{ddkappa3}
\\
D^2\lambda&=&
{2D\kappa\cdot D\lambda\over\kappa},
\label{ddlambda3}
\end{eqnarray}
and the three-dimensional Einstein equation~(\ref{3Ricci}) gives
\begin{eqnarray}
{}^{(\gamma)}R_{mn}=
{3C_1{}^2\over 2}\phi_{,m}\phi_{,n}
+{\kappa_{,m}\kappa_{,n}+\epsilon\lambda_{,m}\lambda_{,n}
\over 2\kappa^2},
\label{3Ricci3}
\end{eqnarray}
where $\epsilon=u/|u|$.
Hence the system is characterized by by 
Eqs.~(\ref{harmonic}),~(\ref{ddkappa3}),~(\ref{ddlambda3}) and 
(\ref{3Ricci3}).
If we choose $\epsilon=+1$ ($u>0$), then Eqs.~(\ref{ddkappa3}) and (\ref{ddlambda3}) 
are combined into a single complex equation
\begin{equation}
D^2\Gamma={2D\Gamma\cdot D\Gamma\over \Gamma+\Gamma^*},
\label{ernst}
\end{equation}
where 
\begin{equation}
\Gamma=\kappa+i\lambda.
\end{equation}
The equation~(\ref{ernst}) is just the same form as the vacuum Ernst equation\cite{Ernst}.
The three-dimensional Einstein equation
\begin{equation}
{}^{(\gamma)}R_{mn}=
{3C_1{}^2\over 2}\phi_{,m}\phi_{,n}
+{2\Gamma_{,(m}^{}\Gamma_{,n)}^*\over (\Gamma+\Gamma^*)^2}.
\label{3Ricciernst}
\end{equation}
The Equations~(\ref{ddkappa3}), (\ref{ddlambda3}) and (\ref{3Ricci3}) are invariant under
\begin{equation}
\phi\mapsto \phi +s_1,
\label{s1}
\end{equation}
and under the {\em Geroch transformation}\cite{Geroch}
\begin{eqnarray}
&&\kappa\mapsto{\kappa\over1-2\lambda s_2+(\lambda^2+\epsilon \kappa^2)s_2{}^2},
\label{s21}\\
&&\lambda\mapsto{\lambda-(\lambda^2+\epsilon \kappa^2)s_2\over1-2\lambda s_2+(\lambda^2+\epsilon\kappa^2)s_2{}^2},
\label{s22}
\end{eqnarray}
in terms of real parameters $s_1$ and  $s_2$. 
In other words, once we have a solution $\phi$ and $\Gamma$, 
then the r.h.s. of Eqs.~(\ref{s1}), (\ref{s21}) and (\ref{s22}) 
give a sequence of solutions parameterized by $s_1$ and $s_2$.
\subsection{Solution determined by a harmonic function}
We here further assume that the potentials $\kappa$ and $\lambda$ are 
also functions of $\phi$;
$\kappa=\kappa(\phi)$, $\lambda=\lambda(\phi)$.
We set $C_1=C_3=i$, $C_2=C_4=0$ and $\epsilon_4=+1$.
Then, $u$ and $\phi$ become
\begin{eqnarray}
u&=&-\sec^2\phi,
\label{iwpu}\\
\Phi&=&-\sqrt{3\over 2}\tan\phi.
\end{eqnarray}
Since $u<0$, we set $\epsilon=-1$.
Then Eqs.~(\ref{ddkappa3}),(\ref{ddlambda3}) and (\ref{3Ricci3}) become
\begin{eqnarray}
\kappa''&=&{(\kappa')^2+(\lambda')^2\over\kappa},
\label{kappaconf}\\
\lambda''&=&{2\kappa'\lambda'\over\kappa},
\label{lambdaconf}\\
{}^{(\gamma)}R_{mn}&
=&\left[-{3\over 2}+{(\kappa')^2-(\lambda')^2\over2\kappa^2}\right]\phi_{,m}\phi_{,n}.
\label{3Ricciconf}
\end{eqnarray}
We further assume that the three-metric is flat: $\gamma_{ij}=\delta_{ij}$, which implies that
${}^{(\gamma)}R_{mn}=0$. Then, the general solution to 
Eqs.~(\ref{kappaconf}), (\ref{lambdaconf}) and (\ref{3Ricciconf}) is
given by
\begin{eqnarray}
\kappa&=&{\rm csch}(\sqrt{3}\phi),\\
\lambda&=&-\coth(\sqrt{3}\phi),
\end{eqnarray}
where the integration constants have been eliminated by redefinition of $\phi$
and the coordinate scaling.
This solution gives the metric
\begin{equation}
g= {\rm sinh}(\sqrt{3}\phi)\cos\phi[(dx^1)^2+(dx^2)^2+(dx^3)^2]
-{(dx^4)^2\over \cos^2\phi} +{\cos\phi\over {\rm sinh}(\sqrt{3}\phi)}(dx^5+w^5{}_i dx^i)^2,
\end{equation}
where $w^5{}_i$ is determined by
\begin{equation}
2\partial_{[i}w^5{}_{j]}
=\sqrt{3}\epsilon_{ijk}\phi_{,k},
\end{equation}
and the Maxwell field strength
\begin{equation}
F_{4i}=\sqrt{3\over 2}\phi_{,i}\sec^2\phi.
\end{equation}
Thus, the solution is determined by an axisymmetric
 harmonic function $\phi$ on flat three-space.

\subsection{Weyl-Papapetrou--type solutions in five dimensions}
To solve Eqs~(\ref{ddkappa3}), (\ref{ddlambda3}) and (\ref{3Ricci3}), we
seek for a solution in the form of Eq.~(\ref{weyl-papapetrou}).
Specifically, we assume the existence of another Killing vector $\xi_3=\partial_\eta$ and
put the three-dimensional metric in the form
\begin{eqnarray}
\gamma=e^{2\sigma(\rho,z)}(d\rho^2+dz^2)+\epsilon_3\rho^2d\eta^2,
\end{eqnarray}
where $\epsilon_3=-f/|f|$.
In this case, the differential operator $D$ in Eqs.~(\ref{ddkappa3}) and
(\ref{ddlambda3}) is regarded as that for the axisymmetric field in
flat three-space as already mentioned.
If we choose $\epsilon=+1$ ($u>0$), then Eq.~(\ref{ddkappa3}) and (\ref{ddlambda3})
give a vacuum Ernst equation (\ref{ernst}). 
The metric function $\sigma$ is determined by integrating
\begin{eqnarray}
{2\over\rho}\sigma_{,\rho}
&=&
{3C_1^2[(\phi_{,\rho})^2-(\phi_{,z})^2]\over 2}
+{(\kappa_{,\rho})^2-(\kappa_{,z})^2
+\epsilon[(\lambda_{,\rho})^2-(\lambda_{,z})^2]
\over2\kappa^2}
,\label{5sigmarho}\\
{1\over\rho}\sigma_{,z}&=&
{3C_1^2\phi_{,\rho}\phi_{,z}\over 2}
+{\kappa_{,\rho}\kappa_{,z}
+\epsilon\lambda_{,\rho}\lambda_{,z}
\over2\kappa^2}.
\label{5sigmaz}
\end{eqnarray}
Therefore, we can generate five-dimensional
Einstein-Maxwell fields
from four-dimensional stationary and axisymmetric vacuum metrics such as
Kerr\cite{Ernst} or Tomimatsu-Sato\cite{TS1} solution.
Here, we show some physically relevant metrics in another way.
In the usual four-dimensional formulation, the following prolate spheroidal coordinate system
$(\tilde x,\tilde y)$ is useful.
\begin{eqnarray}
\rho&=&L_0\sqrt{\tilde x^2-1}\sqrt{1-\tilde y^2},\\
z&=&L_0\tilde x\tilde y,
\end{eqnarray}
where $L_0$ is a constant length scale. Then, the three-metric takes the form
\begin{eqnarray}
\gamma&=&L_0^2\left[e^{2\sigma}(\tilde x^2-\tilde y^2)
\left({d\tilde x^2\over \tilde x^2-1}+{d\tilde y^2\over 1-\tilde y^2}\right)
+\epsilon_3(\tilde x^2-1)(1-\tilde y^2)d\eta^2\right].
\end{eqnarray}
In particular, the Ernst potential for the Kerr metric\cite{Ernst} has a very simple form
in this coordinate system.
In five dimensions, another coordinate system $(x,y)$ defined by
\begin{eqnarray}
\rho&=&{L_0\over 4}xy\sqrt{x^2-2}\sqrt{2-y^2},
\label{qsrho}\\
z&=&{L_0\over 4}xy
\label{qsz}
\end{eqnarray}
turns out to be useful. Then, the three-metric becomes
\begin{eqnarray}
\gamma&=&L_0^2\left[e^{2\sigma}{(x^2-y^2)(x^2+y^2-2)\over 4}
\left({dx^2\over x^2-2}+{dy^2\over 2-y^2}\right)
+\epsilon_3{x^2y^2\over 16}(x^2-2)(2-y^2)d\eta^2\right].
\end{eqnarray}
Let us first solve the Laplace equation $D^2\phi=0$ on this three-space
\begin{eqnarray}
x^{-1}
\partial_x\left[x(x^2-2) \partial_x\phi\right]
+y^{-1}\partial_y\left[y(2-y^2) \partial_y\phi\right]
=0.
\end{eqnarray}
If we assume the solution in the form $\phi=\phi_1(x)+\phi_2(y)$, then
it becomes
\begin{eqnarray}
\phi=(C_5 +C_6)\ln |x|+(C_5 -C_6)\ln \sqrt{x^2-2}
+(C_5 +C_7)\ln|y|+(C_5 -C_7)\ln\sqrt{2-y^2},
\end{eqnarray}
where $C_5 ,C_6$ and $C_7$ are real parameters.
Next, assuming $\epsilon=+1$ ($u>0$), we consider the Ernst equation~(\ref{ernst}).
In the present case, it takes a form
\begin{eqnarray}
x^{-1}\partial_x[x(x^2-2)\partial_x\Gamma]+
y^{-1}\partial_y[y(2-y^2)\partial_y\Gamma]
=
{2[(x^2-2)(\partial_x\Gamma)^2+(2-y^2)(\partial_y\Gamma)^2]\over\Gamma+\Gamma^*}.
\label{ernstspheroidal}
\end{eqnarray}
The simplest solution of Eq.~(\ref{ernstspheroidal}) in rational expression will be
\begin{equation}
\Gamma={2(q^2+1)y\over x+iq y}-xy,
\end{equation}
where $q$ is a real parameter.
Then, the metric becomes
\begin{eqnarray}
g&=&
{L_0^2(x^2+q^2 y^2)
[
e^{2C_2}
|x|^{2C_5 +2C_6}
(x^2-2)^{C_5 -C_6}
|y|^{2C_5 +2C_7} 
(2-y^2)^{C_5 -C_7}
+1]
\over
e^{C_2}
xy[(x^2-2)-q^2 (2-y^2)]
|x|^{C_5 +C_6}
(x^2-2)^{(C_5 -C_6)/2}
|y|^{C_5 +C_7}
(2-y^2)^{(C_5 -C_7)/2}
}
\nonumber\\
&&\times
\left[e^{2\sigma}{(x^2-y^2)(x^2+y^2-2)\over 4}
\left({dx^2\over x^2-2}+{dy^2\over 2-y^2}\right)
+{x^2y^2\over 16}(x^2-2)(2-y^2)d\eta^2\right]\nonumber\\
&&{}+{
e^{2C_2}
|x|^{2C_5 +2C_6}
(x^2-2)^{C_5 -C_6}
|y|^{2C_5 +2C_7} 
(2-y^2)^{C_5 -C_7}
\over
[e^{2C_2}
|x|^{2C_5 +2C_6)}
(x^2-2)^{C_5 -C_6}
|y|^{2C_5 +2C_7} 
(2-y^2)^{C_5 -C_7}
+1]^2}(dx^4)^2\nonumber\\
&&{}-
{[e^{2C_2}
|x|^{2C_5 +2C_6}
(x^2-2)^{C_5 -C_6}
|y|^{2C_5 +2C_7} 
(2-y^2)^{C_5 -C_7}
+1]
\over
e^{C_2}
|x|^{C_5 +C_6}
(x^2-2)^{(C_5 -C_6)/2}
|y|^{C_5 +C_7} 
(2-y^2)^{(C_5 -C_7)/2}}
\nonumber\\
&&\times
{xy[(x^2-2)-q^2 (2-y^2)]\over x^2+q^2 y^2}
\left\{dx^5-{L_0 q(q^2+1)(2-y^2)\over 2[x^2-2-q^2(2-y^2)]}d\eta\right\}^2,
\label{5wpmetric}
\end{eqnarray}
where we set $\epsilon_3=\epsilon_4=+1$, $C_1=1$ and $C_3=1/2$.
The metric function $\sigma$ is determined by
\begin{eqnarray}
\sigma_{,x}&=&{xy(2-y^2)\over 4(x^2-y^2)(x^2+y^2-2)}
\left\{
(x^2-1)y
\left[(x^2-2){}^{(\gamma)}R_{xx}-(2-y^2){}^{(\gamma)}R_{yy}\right]
\right.\nonumber\\
&&{}\left.
+2x(x^2-2)(1-y^2){}^{(\gamma)}R_{xy}
\right\},
\label{sigmax}\\
\sigma_{,y}&=&{xy(x^2-2)\over 4(x^2-y^2)(x^2+y^2-2)}
\left\{
x(y^2-1)
\left[(x^2-2){}^{(\gamma)}R_{xx}-(2-y^2){}^{(\gamma)}R_{yy}\right]
\right.\nonumber\\
&&{}\left.
+2y(x^2-1)(2-y^2){}^{(\gamma)}R_{xy}
\right\}.\
\label{sigmay}
\end{eqnarray}
Using Eq.~(\ref{3Ricciernst}), where 
the three-dimensional Ricci tensor is written in terms of $\phi$ and $\Gamma$,
the integral of Eqs.~({\ref{sigmax}) and ({\ref{sigmay}) is obtained:
\begin{eqnarray}
e^{2\sigma}
&=&
|x^2-2-q^2(2-y^2)|
|x|^{3(C_5 +C_6)^2/2+1/2}
|y|^{3(C_5 +C_7)^2/2+1/2}
(x^2-y^2)^{-3(C_6-C_7)^2/4-1}
\nonumber\\
&&\times
(x^2-2)^{3(C_5 -C_6)^2/4}
(2-y^2)^{3(C_5 -C_7)^2/4}
(x^2+y^2-2)^{-3(C_6+C_7)^2/4+1/4}.
\label{5wpsigma}
\end{eqnarray}
The Maxwell field strength becomes
\begin{eqnarray}
F_{4x}&=&-2\sqrt{6}
\left[
{C_5(x^2-1)-C_6
\over
|x|(x^2-2)}
\right]
\nonumber\\
&&\times
{e^{2C_2}
|x|^{2C_5+2C_6}|y|^{2C_5+2C_7}
(x^2-2)^{C_5-C_6}(2-y^2)^{C_5-C_7}
\over
[e^{2C_2}
|x|^{2C_5+2C_6}|y|^{2C_5+2C_7}
(x^2-2)^{2C_5-2C_6}(2-y^2)^{C_5-C_7}+1]^2},
\label{5wpmax1}\\
F_{4y}&=&-2\sqrt{6}
\left[
{C_5(1-y^2)+C_7
\over
|y|(2-y^2)}
\right]
\nonumber\\
&&\times
{e^{2C_2}
|x|^{2C_5+2C_6}|y|^{2C_5+2C_7}
(x^2-2)^{C_5-C_6}(2-y^2)^{C_5-C_6}
\over
[e^{2C_2}
|x|^{2C_5+2C_6}|y|^{2C_5+2C_7}
(x^2-2)^{2C_5-2C_6}(2-y^2)^{C_5-C_7}+1]^2}.
\label{5wpmax2}
\end{eqnarray}
This solution includes the five-dimensional Myers-Perry solution\cite{MP}
and its electro-magnetic generalization
as special cases.
To see this, we introduce new coordinates
\begin{eqnarray}
r&=&2^{-1/2}r_0x,~~~
\cos\vartheta=2^{-1/2}y,\nonumber\\
\varphi&=&2^{-1/2}\eta,~~~
\psi=2e^{C_2}x^4/r_0,~~~
t=e^{-C_2}x^5,
\end{eqnarray}
and set the parameters
\begin{eqnarray}
L_0&=&2^{1/2}Br_0,\\
e^{C_2}&=&Br_0/2,\\
C_5&=&C_6=C_7=1/2,\\
q&=&a/r_0.
\end{eqnarray}
Then, the metric (\ref{5wpmetric}) reduces to
\begin{eqnarray}
g&=&
\left(1+B^2r^2\cos^2\vartheta\right)
\left[
-{\Delta-a^2\sin^2\vartheta\over \Sigma}dt^2
+
{2a(r^2+a^2-\Delta)\sin^2\vartheta\over\Sigma} dtd\varphi
\right.
\nonumber\\
&&{}
+{(r^2+a^2)^2-a^2\Delta\sin^2\vartheta\over\Sigma}\sin^2\vartheta d\varphi^2
\left.+{\Sigma \over\Delta}dr^2+\Sigma d\vartheta^2\right]
\nonumber\\
&&{}+
\left(1+B^2r^2\cos^2\vartheta\right)^{-2}
r^2\cos^2\vartheta d\psi^2,
\label{melvin-kerrmetric}
\end{eqnarray}
and the field strength given by Eqs.~(\ref{5wpmax1}) and (\ref{5wpmax2}) 
becomes
\begin{eqnarray}
F&=&
{\sqrt{6}B
r^2\cos\vartheta
\over
(1+B^2r^2\cos^2\vartheta)^2}
d\psi\wedge \left(\sin\vartheta d\vartheta-{\cos\vartheta dr\over r}\right).
\label{melvin-kerrmaxwell}
\end{eqnarray}
The metric functions $\Delta$ and $\Sigma$ are given by
\begin{eqnarray}
\Delta&=&r^2-r_0^2,\\
\Sigma&=&r^2+a^2\cos^2\vartheta.
\end{eqnarray}
The vacuum limit $B\rightarrow 0$ gives $F\rightarrow 0$, and
the metric is that of the Myers-Perry solution with a single 
nonzero angular momentum parameter.
On the other hand, when $r_0=0$ and $a=0$, 
Eqs.~(\ref{melvin-kerrmetric}) and (\ref{melvin-kerrmaxwell}) give
the metric
\begin{eqnarray}
g&=&
\left(1+B^2\varrho^2\right)
\left(
-dt^2+d\varrho^2+dz^2
+z^2 d\varphi^2
\right)
\nonumber\\
&&{}+
\left(1+B^2\varrho^2\right)^{-2}
\varrho^2 d\psi^2,
\label{melvin-metric}
\end{eqnarray}
and the Maxwell field
\begin{eqnarray}
F&=&
{\sqrt{6}B\varrho
\over
(1+B^2\varrho^2)^2}
d\varrho\wedge  d\psi,
\label{melvin-maxwell}
\end{eqnarray}
where the cylindrical coordinates
\begin{eqnarray}
\varrho=r\cos\vartheta,~~~
z=r\sin\vartheta,
\end{eqnarray}
have been defined.
This can be thought of as the five-dimensional generalization
of the Melvin's uniform magnetic universe\cite{Melvin}, which in this case describes the 
gravitational field of the planer distribution
of the magnetic flux around ($\varrho=0$).
Thus, Eqs.~(\ref{melvin-kerrmetric}) and (\ref{melvin-kerrmaxwell})
can be interpreted as describing the five-dimensional rotating black hole immersed in
the external magnetic field.
Though its physical meaning has not yet been fully investigated,
the six-parameter solution given by Eqs.~(\ref{5wpmetric}), (\ref{5wpmax1}) and (\ref{5wpmax2})
represents more general Einstein-Maxwell fields.

\section{Einstein-Maxwell fields in arbitrary dimensions}
\label{V}
\subsection{Electro-magnetic generalization of Emparan-Reall static solution}
We here consider the $D$-dimensional Einstein-Maxwell system with $(D-2)$
Killing vector fields.
The simplifying assumption here is $f_{IJ}=\epsilon_I e^{2W_I}\delta_{IJ}$ ($\epsilon_I=\pm1$)
and $\lambda_I=\Phi_I=0$.
Then, the metric has the form
\begin{eqnarray}
g=e^{-2W}\left[e^{2\sigma}(d\rho^2+dz^2)+\epsilon_3\rho^2 d\eta^2\right]
+\sum_{I=4}^D\epsilon_Ie^{2W_I}(dx^I)^2,
\end{eqnarray}
where $\epsilon_3=-\Pi_I\epsilon_I$ and $W=\sum_I W_I$.
Let $\Psi$ be given by
\begin{eqnarray}
\Psi_{,\rho}&=&{\rho\over L_0}\phi_{,z},\\
\Psi_{,z}&=&-{\rho\over L_0}\phi_{,\rho},
\end{eqnarray}
where $L_0$ is a constant length scale.
Then, the integrability condition $\partial_{[p}\partial_{q]}\Psi=0$ requires
 $\phi$ to give a axisymmetric harmonic function on flat three-space
\begin{equation}
D^2\phi=0.
\end{equation}
The Einstein-Maxwell equations (\ref{laplacianlambda}) and (\ref{laplacianphi}) 
are automatically satisfied and Eqs.~(\ref{laplacianf})
become
\begin{eqnarray}
W_{I,\rho\rho}+W_{I,\rho}/\rho+W_{I,zz}&=&
\epsilon_3{(D-3)\rho^2e^{-2W}\over(D-2) L_0^2}
[(\phi_{,\rho})^2+(\phi_{,z})^2],
\label{WI}\\
(W_{,\rho}-1/\rho) \phi_{,z}
-W_{,z} \phi_{,\rho}&=&0.\label{psi}
\end{eqnarray}
Eq.~(\ref{psi}) have simple solutions
\begin{equation}
W=\ln(\rho/L_0)+h(\phi),
\end{equation}
in terms of an arbitrary function $h$ of $\phi$.
This function $h$ is determined by the following equation obtained
from Eq.~(\ref{WI})
\begin{eqnarray}
W_{,\rho\rho}+W_{,\rho}/\rho+W_{,zz}&=&
\epsilon_3 {D-3\over D-2}{\rho^2e^{-2W}\over L_0^2}[(\phi_{,\rho})^2+(\phi_{,z})^2],
\label{W}
\end{eqnarray}
which gives
\begin{equation}
h''=
\epsilon_3 {D-3\over D-2}e^{-2h}.
\end{equation}
The general solution is
\begin{eqnarray}
h=\ln\left[\sqrt{D-3\over D-2}{e^{C_8\phi+C_9}+\epsilon_3 e^{-C_8\phi-C_9}\over 2C_8}\right]
\end{eqnarray}
If we  wright $W_I$ in the form
\begin{equation}
W_I={h(\phi)\over D-3}+\phi_I,
\end{equation}
then, Eq.~(\ref{WI}) gives
\begin{eqnarray}
D^2\phi_I&=&0,\\
\sum_I\phi_I&=&\ln(\rho/L_0).
\end{eqnarray}
Thus, the solution is determined by $(D-3)$ harmonic functions $\phi,\phi_4,\cdots,\phi_{D-1}$.
The metric becomes
\begin{eqnarray}
g&=&{L_0^2\over \rho^2}
\left(e^{{1\over 2}\sqrt{D-3\over D-2}\phi}
+\epsilon_3 e^{-{1\over 2}\sqrt{D-3\over D-2}\phi}
\right)^{-2}
\left[e^{2\sigma}(d\rho^2+dz^2)+\epsilon_3 \rho^2 d\eta^2\right]\nonumber\\
&&{}+\left(
e^{{1\over 2}\sqrt{D-3\over D-2}\phi}
+\epsilon_3 e^{-{1\over 2}\sqrt{D-3\over D-2}\phi}
\right)^{{2\over D-3}}\sum_{I=4}^D\epsilon_Ie^{2\phi_I}(dx^i)^2,
\label{Dstaticmetric}
\end{eqnarray}
where we set $C_8=\sqrt{(D-3)/(D-2)}/2$, $C_9=0$ without loss of generality.
The metric function
$\sigma$ is determined by
\begin{eqnarray}
\sigma_{,\rho}&=&{\rho\over 2}\left[
e^{-2W}(\Psi_{,\rho}^2-\Psi_{,z}^2)+W_{,\rho}^2-W_{,z}^2+\sum_i(W_{i,\rho}^2-W_{i,z}^2)\right]
\nonumber\\
&=&
{1\over 2\rho}
+h'\phi_{,\rho}
+[(h')^2-e^{-2h}][(\phi_{,\rho})^2-(\phi_{,z})^2]
\nonumber\\
&&{}+\sum_I\left[
\left(\phi_{I,\rho}+{h'\over D-3}\phi_{I,\rho}\right)^2
-\left(\phi_{I,z}+{h'\over D-3}\phi_{I,z}\right)^2\right]\\
\sigma_{,z}&=&
\rho\left[\Psi_{,\rho}\Psi_{,z}+W_{,\rho}W_{,z}+\sum_IW_{I,\rho}W_{I,z}\right]
\nonumber\\
&=&
h'\phi_{,z}+\rho\left[{\rho^2\over L_0^2}+(h')^2\right]\phi_{,\rho}\phi_{,z}
\nonumber\\
&&
+\rho\sum_I\left(\phi_{I,\rho}+{h'\over D-3}\phi_{I,\rho}\right)
\left(\phi_{I,z}+{h'\over D-3}\phi_{I,z}\right).
\end{eqnarray}
The Maxwell field strength has the component
\begin{eqnarray}
F_{\eta p}=L_0\left(
e^{{1\over 2}\sqrt{D-3\over D-2}\phi}+\epsilon_3 e^{-{1\over 2}\sqrt{D-3\over D-2}\phi}
\right)^2\phi_{,p},
\end{eqnarray}
for $p=\rho,z$.

The above solution in general represents the gravitational field
of the static charged object.
Here, we give several such examples.
\subsubsection{Five-dimensional Reissner-Nordstroem solution}
The $D$-dimensional Reissner-Nordstroem solution represents
 the hyper-spherically symmetric black hole space-time.
This solution admits $(D-2)$ commuting Killing vector fields
only when $D=4,5$. We consider $D=5$ case here.
The metric of five-dimensional Reissner-Nordstroem solution is given by
\begin{eqnarray}
g&=&
-\left(1-{r_0^2\over r^2}\right)\left(1-{k^2r_0^2\over r^2}\right)
dt^2+
{dr^2\over \left(1-r_0^2/r^2\right)\left(1-k^2r_0^2/ r^2\right)}
\nonumber\\
&&{}+r^2(d\vartheta^2+
\sin^2\vartheta d\varphi^2+\cos^2\vartheta d\psi^2),
\label{5RNmetric}
\end{eqnarray}
where $r_0>0$, $0<k<1$ are constant parameters.
Let us introduce the new coordinates defined by
\begin{eqnarray}
r^2&=&(1-k^2)R^2+k^2r_0{}^2,~~~
t=(1-k^2)^{-1}r_0\eta,~~~\nonumber\\
\psi&=&x^4/r_0,
~~~\varphi=x^5/r_0.
\end{eqnarray}
Then, Eq.~(\ref{5RNmetric}) cab be written as
\begin{eqnarray}
g&=&
\left(1-k^2+{k^2r_0^2\over R^2}\right)
\left({dR^2\over 1-{r_0^2\over R^2}}+R^2d\vartheta^2\right)
-\left(1-k^2+{k^2r_0^2\over R^2}\right)^{-2}
\left(1-{r_0^2\over R^2}\right)d\eta^2\nonumber\\
&&{}+
\left(1-k^2+{k^2r_0^2\over R^2}\right){R^2\over r_0{}^2}
[\sin^2\vartheta (dx^4)^2+\cos^2\vartheta (dx^5)^2]
\nonumber\\
&=&
\left[
\left(1-k^2+{k^2L_0{}^2\over R^2}\right)^2
{R^4\sin^2 2\vartheta\over 4L_0{}^4}\right]^{-1}\nonumber\\
&&\times
\left[
\left(1-k^2+{k^2L_0{}^2\over R^2}\right)^3
{R^4\sin^2 2\vartheta\over 4L_0^4}
\left({dR^2\over 1-{r_0^2\over R^2}}+R^2d\vartheta^2\right)
-{1\over 4}\left(1-{r_0{}^2\over R^2}\right){R^4\sin^2 2\vartheta\over r_0{}^2}d\eta^2
\right]
\nonumber\\
&&+\left(1-k^2+{k^2r_0{}^2\over R^2}\right){R^2\over L_0{}^2}
\left[\cos^2\vartheta (dx^4)^2
+\sin^2\vartheta (dx^5)^2\right].
\end{eqnarray}
One can determine the canonical coordinates $(\rho,z)$
from this form as
\begin{eqnarray}
\rho&=&{1\over 2}\left(1-{r_0{}^2\over R^2}\right)^{1/2}{R^2\sin2\vartheta\over r_0},\\
z&=&{1\over 2r_0}\left(1-{r_0{}^2\over 2R^2}\right)R^2\cos 2\vartheta.
\end{eqnarray}
Then, the metric is written in the form of Eq.~(\ref{Dstaticmetric})
\begin{eqnarray}
g&=&{r_0^2\over \rho^2}
\left(e^{\phi/\sqrt{6}}
-e^{-\phi/\sqrt{6}}
\right)^{-2}
\left[e^{2\sigma}(d\rho^2+dz^2)-\rho^2 d\eta^2\right]\nonumber\\
&&{}+\left(
e^{\phi/\sqrt{6}}
-e^{-\phi/\sqrt{6}}
\right)\left[e^{2\phi_4}(dx^4)^2+e^{2\phi_5}(dx^5)^2\right].
\end{eqnarray}
The potentials $\phi$, $\phi_4$ and $\phi_5$, and the metric function
$\sigma$ are given by
\begin{eqnarray}
\phi_4
&=&
{1\over 4}\ln{\nu^2(\mu+\nu)\over \mu+\nu-1},
\\
\phi_5
&=&
{1\over 4}\ln{\mu^2(\mu+\nu)\over \mu+\nu-1},
\\
{\phi\over\sqrt {6}}&=&\ln\left\{
{(1-k^2)(\mu+\nu)-1\over2(\mu+\nu-1)}
+\sqrt{
\left[
(1-k^2)(\mu+\nu)-1\over2(\mu+\nu-1)
\right]^2+{\mu+\nu\over\mu+\nu-1}
}\right\}\nonumber\\
&&{}
+{1\over 2}\ln{\mu+\nu-1\over\mu+\nu},\\
e^{2\sigma}
&=&
\left[(1-k^2)(\mu+\nu)-1\over2(\mu+\nu-1)\right]^3
{
\mu\nu(\mu+\nu)^3
\over
(\mu+\nu-1/2)^3(\mu+\nu+1/2)+(\mu-1/4)(\nu-1/4)
},
\end{eqnarray}
where 
\begin{eqnarray}
\mu&=&
{\sqrt{\rho^2+(z-r_0/2)^2}-z+r_0/4\over r_0},\\
\nu&=&{\sqrt{\rho^2+(z+r_0/2)^2}+z+r_0/4\over r_0},
\end{eqnarray}
have been defined.
The Maxwell field is determined by the potential
\begin{equation}
\psi={3\over 2}{k(\mu+\nu-1/2)\over
(1-k^2)(\mu+\nu-1/2)+1}.
\end{equation}
\subsubsection{Charged black ring solution}
Finally, we briefly discuss the charged black ring solution in five dimensions.
In five or higher-dimensional space-time, 
the black hole need not have spherical topology\cite{Cai-Galloway}.
In fact, a vacuum metric representing the rotating black ring
with $S^2\times S^1$ horizon
has been provided by Emparan and Reall\cite{BR}.
Elvang has also obtained the supergravity generalization of the black ring solution\cite{CBR}.
There is also the static vacuum black ring solution in the Emparan-Reall static 
class\cite{Dstatic}, which is not completely regular due to the
disk-like deficit to prevent the black ring from collapsing.
Here we just give a charged generalization of the static black ring solution.
The metric of the static charged black ring is given by
\begin{eqnarray}
g&=&-{(1-\mu x)(1-\mu y)\over[(1-\mu y)-k^2 (1-\mu x)]^2}dt^2
+{1-\mu y-k^2(1-\mu x)\over A^2(x-y)^2 }
\left[{1-\mu y\over 1-x^2} dx^2
+{1-\mu x\over y^2 -1}dy^2 \right.
\nonumber\\
&&{}\left.+ 
{(1-\mu x)(y^2-1)\over 1-\mu y}d\psi^2 
+{(1-\mu y)(1-x^2)\over 1-\mu x}d\varphi^2\right],
\end{eqnarray}
with the Maxwell field 
\begin{eqnarray}
F&=&-\sqrt{3\over2}{kdt\wedge (dx -dy)\over r_0[1-\mu y-k^2(1-\mu x)]^2},
\end{eqnarray}
where $-1< k<1$ and $A>0$ are constant parameters. This solution has a vacuum limit $k\rightarrow 0$,
which corresponds to the Emparan-Reall static black ring solution.
This solution represents a charged black ring with the event horizon homeomorphic to
$S^2\times S^1$. One might expect that the Coulomb repulsion force can prevent the
black ring from collapsing, though it is not the case as far as this solution is concerned.
As in the neutral case, there is always a conical singularity
either inside or outside the black ring
at the symmetric axis where $g_{\varphi\varphi}=0$.

\section{Summary}
\label{VI}
We have considered the $D$-dimensional Einstein-Maxwell fields admitting
$(D-3)$ Killing vector fields which commutes with each other.
The Maxwell fields can be described by $(D-2)$ electric and magnetic scalar
potentials.
The Einstein equations assure the existence of the twist potentials, then the 
Einstein-Maxwell fields are completely determined by $(D-2)(D-3)/2$ gravitational potentials
$f_{IJ}$, $(D-3)$ twist potentials $\lambda_I$, $(D-3)$ electric potentials
$\Phi_I$, the magnetic potential $\Psi$  and the three-metric $\gamma_{\mu\nu}$,
reducing  to a harmonic map coupled to gravity on the three-dimensional
base space.
If we further assume the existence of another Killing vector field which commutes with the
others, and the integrability condition of the two-dimensional distribution orthogonal
to the all Killing vector fields, then the system is described by the harmonic map coupled to gravity 
on two-dimensional base space, which is the higher-dimensional generalization of the
Weyl-Papapetrou class in the four-dimensional Einstein-Maxwell system.

We have found some explicit solutions to the system.
We show that there is a five-dimensional solution determined by a single harmonic function
which is axisymmetric in the three-dimensional flat space.
There is also a solution determined by a harmonic function and a solution to the
vacuum Ernst equation. We have directly solve these equations in a quasi-spheroidal 
coordinate system given by Eqs.~(\ref{qsrho}) and (\ref{qsz}), and have 
obtained as a special
case a rotating black hole immersed in a magnetic universe.

We have also studied the electro-magnetic generalization of the Emparan-Reall
static solutions. We show that the solution is determined by $(D-3)$ axisymmetric harmonic functions in the 
three-dimensional flat space.
We have also give the five-dimensional charged black ring solution which has a conical
singularity at the symmetric axis.

The system considered here will be useful in extending the black hole uniqueness theorem
to the higher dimensional cases.
We hope that the charged generalization of the Myers-Perry black hole is obtained
through further study of this system.

\section*{Acknowledgments}
We are grateful to Yoshiyuki Morisawa and Kin-ya Oda for discussions.
D.I. was supported by JSPS Research, and this research was supported in part by the
Grant-in-Aid for Scientific Research Fund (No.6499).

\end{document}